\documentclass[12pt,twoside]{article}

\usepackage{authblk}
\usepackage{amssymb,amsmath}
\usepackage{graphicx}
\usepackage{epsfig}
\usepackage{lscape}
\usepackage{amssymb}

\begin{document}

\title{Quantum Phase Transitions in Quantum Dots}
%\author{I. G. Rau, S. Amasha, Y. Oreg, D. Goldhaber-Gordon}
\author[1]{\normalsize I. G. Rau}
\author[1]{S.Amasha}
\author[2]{Y. Oreg}
\author[1]{D. Goldhaber-Gordon}
\affil[1]{\footnotesize Geballe Laboratory for Advanced Materials, Stanford University, Stanford, CA 94305, USA}
\affil[2]{Department of Condensed Matter Physics,
Weizmann Institute of Science, Rehovot 96100, Israel}
\date{March 2010}
\maketitle

\section{Introduction}
\label{sect:Intro}

Over the past 25 years, developments in nano-fabrication and  low temperature cooling techniques, as well as the development of new calculational tools, have given rise to a new field of research: the design of artificial nanostructures that imitate conventional materials.  Initially these nanostructures could only mirror the behavior of some basic building blocks of matter: for example, a quantum dot acts as an artificial atom and can be used to model a localized magnetic moment \cite{Kouwenhoven1997_ShellFilling,glazmanraikh.kondo,nglee.kondo}. More recent technological advances have made it possible to build increasingly complex artificial structures: multi-dot systems coupled to distinct electron reservoirs now offer a way to study the interplay between single spins and conduction electrons, the transport of heat through low dimensional structures, and charge as well as spin transport through atoms and molecules \cite{ron, craig.2ikm, pekola2007_RFSETFridge, Scott2009_MolecKondoScaling}.

The advantage offered by these artificial structures is the fact that their individual properties such as energy spectrum, magnetic moment, coupling to the environment, and the spatial distribution of the wavefunction are tunable in-situ by ``knobs" such as gate voltage, magnetic field, or voltage bias. In real materials these microscopic properties can be changed only by altering a material's structure or chemical composition, or by changing an external parameter such as applied pressure or magnetic field. As a consequence, it is difficult to tune a single microscopic parameter without changing other parameters or even the effective Hamiltonian of the system, and therefore the two phases of a quantum phase transition often belong to two different though related materials. In contrast, quantum dot parameters can be tuned continuously and nearly independently of one another, allowing the exploration of the complete phase diagram, including the quantum critical point (QCP). These artificial structures have advantages and disadvantages that make them complementary to bulk materials: it is difficult to measure thermodynamic quantities such as specific heat or magnetic susceptibility but one can investigate transport properties of impurity quantum phase transitions which will be described in the next paragraphs.

Quantum phase transitions (QPTs) are a class of phase transitions that occur at absolute zero temperature as one varies a parameter other than temperature. Second order (continuous) QPTs are driven by quantum fluctuations of the order parameter, which have properties completely different from those of the familiar thermal fluctuations. The inherent zero-temperature nature of the QPT makes it impossible to observe directly. However, in the low-temperature limit, correlation lengths and times diverge near the transition between two ground states. These long-range correlations influence the behavior of the system at finite temperature: near the QCP, a distinctive set of excitations can be accessed experimentally. These excitations are collective, so that Fermi-liquid theory fails to describe the physics in the quantum critical region. The behavior of the system as a function of external parameters obeys scaling laws with non-trivial exponents that are determined only by the universality class of the transition and not by the microscopic details.

We generally think of second-order phase transitions (classical or quantum) as requiring the thermodynamic limit of system size. However, for a special kind of QPT involving a boundary (e.g. an interface or impurity) embedded in a bulk system, only the degrees of freedom belonging to the boundary become critical, while the thermodynamic limit is only required for the bulk part of the system. Boundary phase transitions show the same fascinating quantum critical behavior as bulk transitions. However, while the entropy at a bulk QCP vanishes at zero temperature, an impurity QCP can have residual entropy: fluctuations are strong enough to preserve some of the local degrees of freedom.

Impurity QPTs have been predicted to occur in various materials, but at first glance one might expect their experimental observation to be impeded by effects present in real materials: the crystalline field anisotropy or exchange or dipolar interaction between spins should break the symmetry of the low-lying states at the QCP. While this generally happens, there are some exceptions where the required symmetries occur naturally, as is the case for f-electron heavy fermion materials based on Uranium or Cerium ~\cite{seaman2ck, fischer2ck, maple.2ck}, Copper or Titanium nano-constrictions ~\cite{ralph2ck.1, upadhyay.2ck}, and glassy metal junctions or single crystals~\cite{keijsers.2ck, cichorek.2ck}. Non-Fermi liquid behavior, as evidenced by non-trivial power laws in thermodynamic and transport quantities, has been measured, but it has proved difficult to connect directly to theory, and alternative explanations have been suggested~\cite{wingreen.2ck}.

Two decades ago it was noticed that a quantum dot with a net spin-1/2, coupled by tunneling to an electron reservoir, has the same Hamiltonian as a magnetic site interacting with electrons in a host metal \cite{glazmanraikh.kondo, nglee.kondo}. This quantum impurity problem is the basis for the Kondo effect, in which the local magnetic moment is fully screened by the spins of the conduction electrons, and the conduction electrons can be described as a Fermi liquid, even near the local site. This review will focus on variations of this system, involving one or more quantum dots, (generally each with spin-1/2) interacting with one or more electron reservoirs. For example, adding an additional reservoir of conduction electrons (screening channel) leads to a ``2-channel Kondo" (2CK) system with the Hamiltonian
\begin{equation}
H_{\rm 2CK}= J_1 \vec{s}_1 \cdot \vec{S} + J_2 \vec{s}_2\cdot \vec{S},
\end{equation}
where $J_1, J_2>0$ are the antiferromagnetic interaction between the spins of the reservoir electrons $\vec{s}_1$ and $\vec{s}_2$ and the local spin $\vec{S}$. When the two reservoirs are symmetrically coupled to the impurity ($J_1=J_2$) the local magnetic moment is overscreened, the system shows local non-Fermi liquid (nFL) behavior, and there is a residual entropy at zero temperature. This model has been used to explain the experimentally observed specific heat anomalies in certain heavy fermion
materials~\cite{maple.2ck, seaman2ck, fischer2ck} as well as transport signatures in metallic nano-constrictions ~\cite{ralph2ck.1, ralph2ck.2}.

\begin{figure}[t]
\setlength{\unitlength}{1cm}
\begin{center}
\begin{picture}(12,3.8)(0,0)
\put(0,0){\includegraphics[width=6cm,keepaspectratio=true]{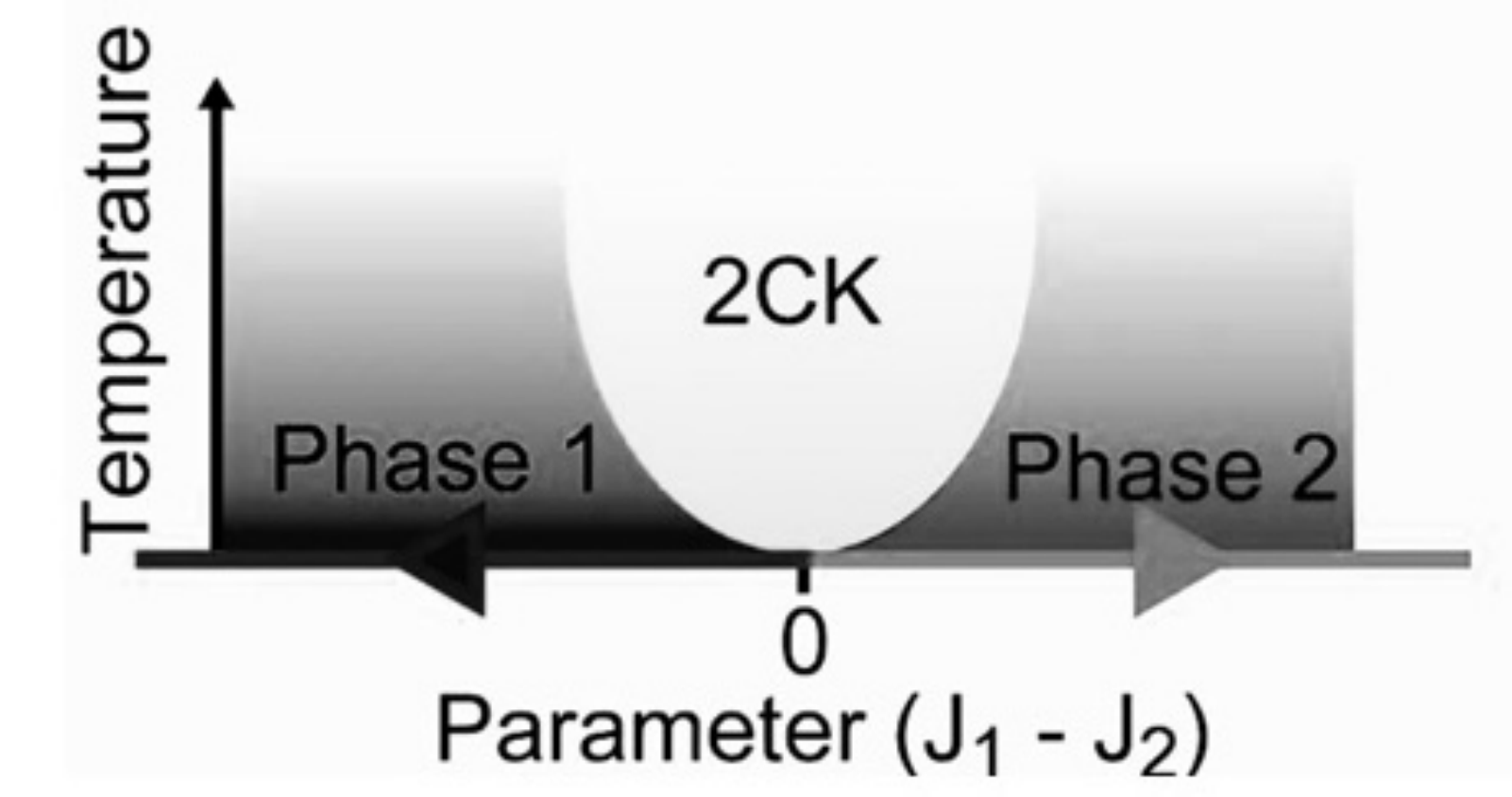}}
\put(0.3,3.5){\makebox(0,0){(a)}}

\put(8,1){\includegraphics[width=4cm,keepaspectratio=true]{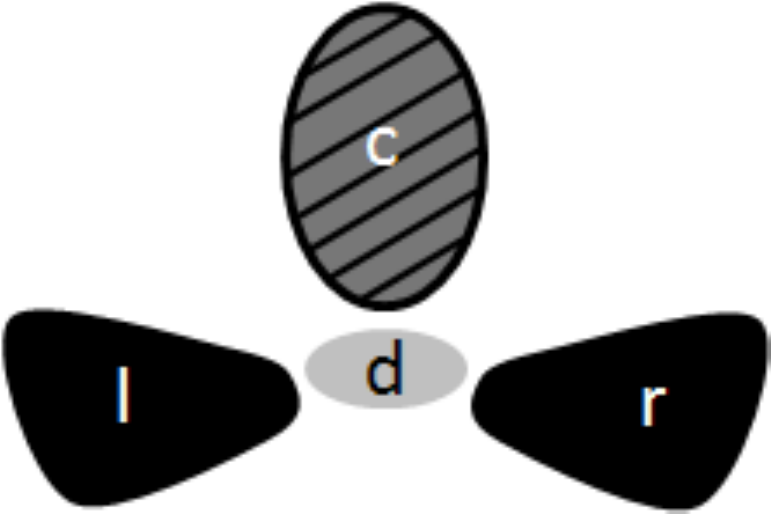}}
\put(8,3.5){\makebox(0,0){(b)}}

\end{picture}
\end{center}
\caption{(a) Schematic phase
diagram of the 2CK system, including the quantum critical point and
the associated parabolic-shaped critical region. The vertical axis
is labeled temperature, but any energy (frequency, bias voltage)
will play a similar role. The tuning parameter shown is the
difference in exchange coupling to two reservoirs, but a similar
phase diagram is expected if the exchange couplings are kept equal
and the tuning parameter is instead magnetic field. (b) Schematic of a quantum system that realizes the 2CK model. A spin on the small dot \textit{d} can influence transport from the left \textit{l} to the right \textit{r} lead. \textit{c} is a finite electron reservoir formed by a large quantum dot which will act as another screening channel for the small dot.
}
\label{fig:2CKwithPetals}
\end{figure}

The two-channel Kondo state also corresponds to the QCP in a boundary QPT between two distinct single-channel Kondo states. This transition takes place as a function of the relative couplings of the two channels to the impurity. For equal coupling, the system is in the 2CK state described above, but when one channel is more strongly coupled the traditional Kondo screening behavior is recovered. Thus the two ground states on either side of the transition are both the standard Kondo singlet state (left and right regions in the phase diagram in Fig.~\ref{fig:2CKwithPetals}(a)), with a different set of electrons participating in the screening of the local moment in each phase. In the quantum critical region (center region in Fig.~\ref{fig:2CKwithPetals}(a)) temperature fluctuations mask the channel anisotropy and the long range correlations of the  2CK ground state dominate the behavior of the system, resulting in nFL scaling laws (see section ~\ref{sec:ExScalingCurves}).

Another interesting model system is that of two spins coupled to a bath of conduction electrons and to each other. This two-impurity Kondo (2IK) system undergoes a QPT from a ground state with total spin 0 (the two impurities antiferromagnetically coupled, forming a spin singlet) to another ground state where the impurities are individually screened by the electrons in the reservoirs. The physics of a 2IK system can also be studied using a singlet-triplet transition in a quantum dot. In this case magnetic field can induce a transition from two spins on the quantum dot in a singlet state to a state where an electron reservoir attempts to screen part or all of the triplet state in the dot.

After briefly reviewing the theoretical predictions for the 2CK model, we will describe the quantum dot system that implements this model, and the observation of the nFL state. We will also discuss other experimental observations of the 2CK model, and the related 2IK model. We conclude with forward-looking theoretical predictions and promising experimental designs.

\section{The Kondo Effect and Quantum Dots: \mbox{theory}}
\subsection{Brief History of the Kondo Effect}

In 1964, motivated by a set of striking experiments~\cite{Haas34,Sarachik64}, Kondo calculated the effect on a metal's resistivity of magnetic impurities which can scatter electrons.  Using perturbation theory on the s-d model
\begin{equation}
\label{eq:s-d}
 H_{s-d}=\sum_k  \epsilon_{k s} \psi^\dagger_{k s}  \psi_{k s}+ J \sum_i \vec \sigma_i \cdot \vec S_i,
\end{equation}
he was able to explain the observed upturn in resistivity at low temperatures.
Here $\vec S_i$ is the $i^{th}$ impurity spin, $\vec \sigma_i$
is the spin of the conduction electrons at the location of the $i^{th}$
impurity, and $\psi$ and $\psi^\dagger$ represent the annihilation and creation operators for the
conduction electrons with given momentum and spin. However, Kondo's perturbative results break down below the Kondo temperature
\begin{equation}
\label{eq:Tk}
T_K = D e^{1/ (J \nu)}
\end{equation}
where D is the conduction electron bandwidth, $J$ the antiferromagnetic coupling of the local spin to the conduction electrons, and $\nu$ the thermodynamic density of states of the conduction electrons. At first glance, this local spin coupled to a non-interacting Fermi sea is a simple system, but below the Kondo temperature the repeated impurity spin-flips and the corresponding response of the Fermi sea lead to complex many-body dynamics. This became known as the Kondo problem and the attempts to find a solution led to the development of a series of useful mathematical and physical concepts such as the Abrikosov-Suhl resonance ~\cite{Abrikosov65,Shul65}, bosonization \cite{Schotte69b}, and renormalization group \cite{Wilson75} as applied to the alternating Coulomb gas \cite{Yuval70,Anderson70}.

Only a few years later Schrieffer and Wolff~\cite{Schriefer66} showed that the s-d model is equivalent to the Anderson model
\begin{equation}
\label{eq:Anderson}
 H_{A} = \sum_k  \epsilon_{k s} \psi^\dagger_{k s}  \psi_{k s}+ U n_{\downarrow} n_{\uparrow}+\sum_s \left[\epsilon_{d s} d^\dagger_s d_s
 + \left(V \psi^\dagger_s(0) d_s + h.c. \right)\right],
\end{equation}
in the limit in which local charge fluctuations can be neglected.
Here $U$ is the interaction parameter on the impurity site, $V$
the hybridization between the conduction electrons and the
impurity and the $d$'s are the creation and annihilation
operators of the impurity with $n_s=d^\dagger_s d_s$.

The generalization of the s-d model to multiple screening channels was introduced in 1980 by Nozieres and Blandin~\cite{Nozieres80}
\begin{equation}
\label{eq:multichannel}
H_{MCK}=\sum_k  \epsilon_{k s \alpha} \psi^\dagger_{k s\alpha }  \psi_{k s \alpha}+ J \sum_\alpha \vec \sigma_\alpha \cdot \vec S.
\end{equation}
In the multichannel Kondo system they identified a non-trivial fixed point associated with a novel metallic phase, unlike in the single channel case, where at energies well below $T_K$ the conduction electrons around the spin-$\frac{1}{2}$ impurity behave as a Fermi liquid. This new phase joined Fermi liquids, Luttinger liquids, fractional quantum Hall systems and disordered systems with Coulomb interaction as the only known classes of metals. In parallel, Zawadowski
studied two-level systems~\cite{Zawadowski80} that display similar behavior.

In the early '90s \cite{AffleckandLudwig90,Affleck93} Affleck and
Ludwig applied boundary conformal field theory
(BCFT) to calculate the properties of the multichannnel Kondo
system at low temperature \cite{Affleck95}. However, it is difficult to use conformal field theory to identify physical situations in which the interesting boundary
state is realized. Instead, one can use numerical renormalization group, which is very reliable in giving
the whole (equilibrium) crossover behavior of quantum
impurity problems, including the single-channel Kondo
effect and the two-channel and the two-impurity Kondo effects with their special points of nFL  behavior. However, it demands special expertise and many computational tricks that
help to accelerate the calculations. Therefore other methods were developed to approximate
simple solutions such as Slave bosons \cite{Kotliar86}, the non-crossing
approximation (NCA) \cite{Cox93}, Equation of Motion (EOM) method~\cite{Krawiec07}, functional renormalization group (FRG)
\cite{Andergassen06}, and the density matrix renormalization
group (DMRG)\cite{White92} and its novel time dependent version
\cite{Dechiara08,Schollwock05}.

\subsection{Theory of Conductance though Quantum Dots}
\label{sec:conductance}

In this section we briefly review theoretical expressions for the conductance through a quantum dot. The possibility of observing the Kondo effect in a quantum dot was established once Anderson-type models were used to describe dots. A convenient way of probing the state of the dot is by measuring its conductance. Using the Anderson Hamiltonian, one can derive scaling relations for this conductance in different regimes (i.e., 1CK and 2CK). In the study of bulk materials, scaling relations for thermodynamic quantities are a powerful tool for probing QPTs and low energy fixed points, and in a similar way the conductance scaling relations presented below will play a crucial role in identifying the 2CK QPT in a system of quantum dots.

We consider the conductance from the left to the right lead of the system in Fig.~\ref{fig:2CKwithPetals}(b)~\cite{dgg.2ck}.
As explained below, with the proper tuning of the system parameters this system exhibits 2CK physics.

The modified Anderson Hamiltonian that describes the system is given by
\begin{eqnarray} \label{eq:modelhamiltonian}
  H &=& \sum_{k\sigma} \varepsilon_{lk \sigma} l_{k\sigma}^\dagger l_{k\sigma}
  +\sum_{k\sigma} \varepsilon_{rk \sigma} r_{k\sigma}^\dagger r_{k\sigma}
  +\sum_{k\sigma} \varepsilon_{ck\sigma}c_{k\sigma}^\dagger  c_{k\sigma} \nonumber \\
  &+& E_c( n - \mathcal{N} )^2
  +\varepsilon_{d \sigma} d_{\sigma}^\dagger d_{\sigma}
  +U n_{d\uparrow}n_{d\downarrow} \nonumber\\
  &+&\sum_{k \sigma} \left(
  t_{k l}
   l_{k\sigma}^\dagger d_{\sigma}+ t_{k r} r_{k\sigma}^\dagger d_{\sigma}+t_{k c} c_{k\sigma}^\dagger
  d_{\sigma}
  + \mathrm{h.c.} \right).
\end{eqnarray}
Here the operators $l_{\sigma k}, (l^{\dagger}_{\sigma k})$, $r_{\sigma k}, (r^{\dagger}_{\sigma k})$, and $c_{\sigma k}, (c^{\dagger}_{\sigma k})$ are the annihilation (creation) operators of a free electron in state $k$ with spin $\sigma$, for the left and right leads and the finite dot, respectively. The fourth term describes the interaction energy $E_{c}$ of the large dot, where $n=\sum_{k\sigma} c^\dagger_{k \sigma} c_{k \sigma}$ is the number operator of the electrons in the large dot, while $\mathcal{N} \propto V_{g}$ gives the effective interaction with an external gate voltage $V_{g}$. The fifth term describes the level spacing of the small dot, with $d_\sigma\,(d^{\dagger}_\sigma)$  the annihilation (creation) of an electron with spin $\sigma$ on the small dot. The sixth term describes the charging energy $U$ of the small dot (with $n_{d\sigma}= d^\dagger_\sigma d_\sigma$). The last term describes the hopping matrix elements $t_{k \alpha}$ between the small dot and the two leads or the large dot (here $\alpha= l, r, c$). We assume that the hopping matrix elements are independent of $k$, and define $t_{\alpha} \equiv  t_{k \alpha}$. In this Hamiltonian, we assume there is only one level in the small dot. This assumption is valid at an energy scale smaller than the level spacing of the small dot, which is the ultraviolet cutoff of the theory. For simplicity we have also ignored the inter-dot electrostatic coupling, which does not qualitatively change the picture.

Typically, in an experiment a bias voltage $V$ is applied between the left and the right leads of the small dot, and the resulting current $I$ is measured. From this measurement, one can extract the differential conductance,
\begin{equation}
\label{def:DifferentialConductanceApp}
G(V,T)=\frac{dI}{dV},
\end{equation}
Using \ref{eq:modelhamiltonian}, one can derive theoretical expressions for $dI/dV$ as a function of $V$, $T$, and the parameters of the Hamiltonian. We briefly outline a calculation here; more details of these calculations are discussed in references \cite{GlazmanPustilnik03} and \cite{Oreg2007_2CKDots}. $I$ is related to the local spectral density of
the small dot by\footnote{Eq.~(\ref{eq:GVT}) is valid only if the coupling matrices to the left and right leads are proportional to one another. When there is only one state in the dot~\cite{Meir92}, as we have assumed in Eq. \ref{eq:modelhamiltonian}, this condition is satisfied.}

\begin{equation}
\label{eq:GVT}
I=\frac{e}{h} \sum_{s=\uparrow,\downarrow}\int d \varepsilon \left[
f_l\left(\varepsilon\right)-f_r\left(\varepsilon\right)\right]
\left[\frac { 2 \Gamma_l \Gamma_r}{\Gamma_l+\Gamma_r}
A_s\left(\varepsilon\right)\right].
\end{equation}
Here $f_{l(r)}$ is the Fermi function in the left (right) lead, $\Gamma_{l(r)}=\pi \nu \left|t_{l(r)}\right|^2$, we assume that $\Gamma_{l(r)}$ does not depend on spin, and we assume that the density of states $\nu$ is identical for the two leads. In this equation, $A_s\left(\varepsilon\right)$ is the spectral density of spin $s$ and it is related to the retarded Green's function of the dot $G^{\rm ret}_{ds}(\varepsilon)$ through
\begin{equation}
\label{eq:spectral_density_and_retarded_green_function}
A_s\left(\varepsilon\right) = -2 \;\; {\rm Im}\left[G^{\rm ret}_{ds}\left(\varepsilon\right) \right].
\end{equation}

In general, $A_s(\varepsilon)$ may depend upon the chemical potential of the left and right leads. However, in the limit where the bias $V$ is small or the tunneling matrix element to one of the leads is much smaller than to the other lead, it is possible to write 

\begin{equation}
\label{eq:GVTsmalltl}
G(V,T) = \nu \tilde{G_0} \sum_{s=\uparrow,\downarrow}\int d \varepsilon
f^\prime\left(\varepsilon-eV \right) {\rm Im}\;{\cal T}^\sigma(\varepsilon),
\end{equation}
where ${\cal T}^\sigma(\varepsilon)$ is the scattering ${\cal T}$-matrix and $\tilde{G_0}$ is a proportionality constant depending on $\Gamma_{l,r}$ \cite{Oreg2007_2CKDots}. As a part of this calculation one sees that because electrons can move from the left to the right lead, these leads do not form independent screening channels, but rather form one effective screening channel. This channel then competes with the second screening channel formed by the large dot. This is also discussed in Sec. \ref{sec:TwoCK_DQD}.

\subsection{Examples of Conductance Scaling Curves}
\label{sec:ExScalingCurves}
Using Eq.~\ref{eq:GVTsmalltl} together with the specific expressions for the scattering matrix ${\cal T}$ of
Ref.~\cite{Affleck93} we can calculate the scaling curves for the 2CK and the 1CK Kondo models (the 1CK model is in fact just the 2CK model with strongly asymmetric channel coupling). These curves will be essential for identifying the QPT in the data in \ref{sec:TwoCK_DQD}.

\subsubsection{$G(V,T)$ in the two-channel Kondo case}

Using the results for the scattering matrix from Ref.~\cite{Affleck93} we can write
\begin{equation}
\label{eq:GVT2cka}
G(V,T)=G_0\frac{1}{2}  \left[1- \sqrt{\frac{\pi T}{T_{K2}}}
F_{2CK}\left(\frac{eV}{ \pi T}\right) \right],
\end{equation}
where the function $F_{2CK}$ is given in Ref. \cite{Oreg2007_2CKDots}. Its asymptotes are

\begin{equation}
\label{eq:F2CKlimits}
F_{2CK}(x) \approx \left\{ \begin{array}{cc}
 c \ x^2+1 &  {\rm{ for }}~ x \ll 1   \\
 \frac{3} {\sqrt{\pi}}  \sqrt{x} & {\rm{ for }} ~ x \gg 1   \\
\end{array} \right.
\end{equation}
where it was numerically determined that $c=0.748336$. Setting $V= 0$ we find
\begin{equation}
\label{eq:G0T2ck1b}
G(0,T)=\frac{1}{2} G_0\left(1- \sqrt{\frac{\pi T}{T_{K2}}} \right) .
\end{equation}

Examining Eqs.~(\ref{eq:GVT2cka}) and (\ref{eq:G0T2ck1b}), we find that the scaling relation for the conductance of a quantum dot in the 2CK regime is

\begin{equation}
\label{eq:ScalingCurves2}
 \frac{2}{G_0} \frac{G(0,T)-G(V,T)}{\sqrt {\pi T /T_{K2}}} = Y\left(\frac{\left|e V\right|}{\pi T}
 \right),
\end{equation}
with the scaling function $Y(x)=F_{2CK}(x)-1$. One of the remarkable features of the 2CK model is that one obtains the full scaling function rather than just a power law approximation valid only at low energies. The scaling curve $Y(x)$ was used
in Ref.~\cite{ron}.

\subsubsection{$G(V,T)$ in the single-channel Kondo case}
The 1CK case can be viewed as a limit of the 2CK case when one of the channels, either the large dot or the leads is more strongly coupled to the small dot than the other channel. In this case, at temperatures smaller then the crossover temperature $T_\Delta$~\cite{pustilnik.2ck} we should have  regular single-channel behavior: the small dot should be Kondo screened by the more strongly coupled channel. Here $\Delta J$ is the asymmetry parameter and is taken to be positive when the leads are more strongly coupled to the small dot.

For $T,e V \ll T_{\Delta}$ we can use the scattering matrix given by~\cite{Affleck93} and obtain, in the limit $\left|t_l\right| \ll \left|t_r\right|$ or vice-versa
\begin{equation}
\label{eq:GVT1ck}
 G(V,T)= G_0\left\{
 \theta(\Delta J)-{\rm{sign}}(\Delta J) \left(\frac{\pi
T}{T_{\Delta}} \right)^2 \left[1+\frac{3}{2}\left(\frac{e V}{\pi
T}\right)^2 \right] \right\}.
\end{equation}

At zero bias, the conductance is given by:
\begin{equation}
\label{eq:G0T1ck}
 G(0,T)= G_0\left\{
 \theta(\Delta J)-{\rm{sign}}(\Delta J) \left(\frac{\pi
T}{T_{\Delta}} \right)^2  \right\}.
\end{equation}

Examining Eq.~(\ref{eq:GVT1ck}) we note that the scaling relation for the conductance of a quantum dot in the 1CK regime is:

\begin{equation}
\label{eq:ScalingCurves}
 \frac{1}{G_0} \frac{G(0,T)-G(V,T)}{\left(\pi T /T_{\Delta}\right)^2} = {\rm{sign}}\left(\Delta J\right)
\frac{3}{2}\left(\frac{e V}{\pi T}\right)^2.
\end{equation}
This scaling curve is different from the 2CK curve, and was used to analyze the data in Ref. \cite{ron}.

\section{The Kondo Effect and Quantum Dots: Experiment}

\subsection{The Two-channel Kondo Effect in a Double Quantum Dot}
\label{sec:TwoCK_DQD}

\begin{figure}[t]
\begin{center}
\resizebox{12cm}{!}{
\includegraphics{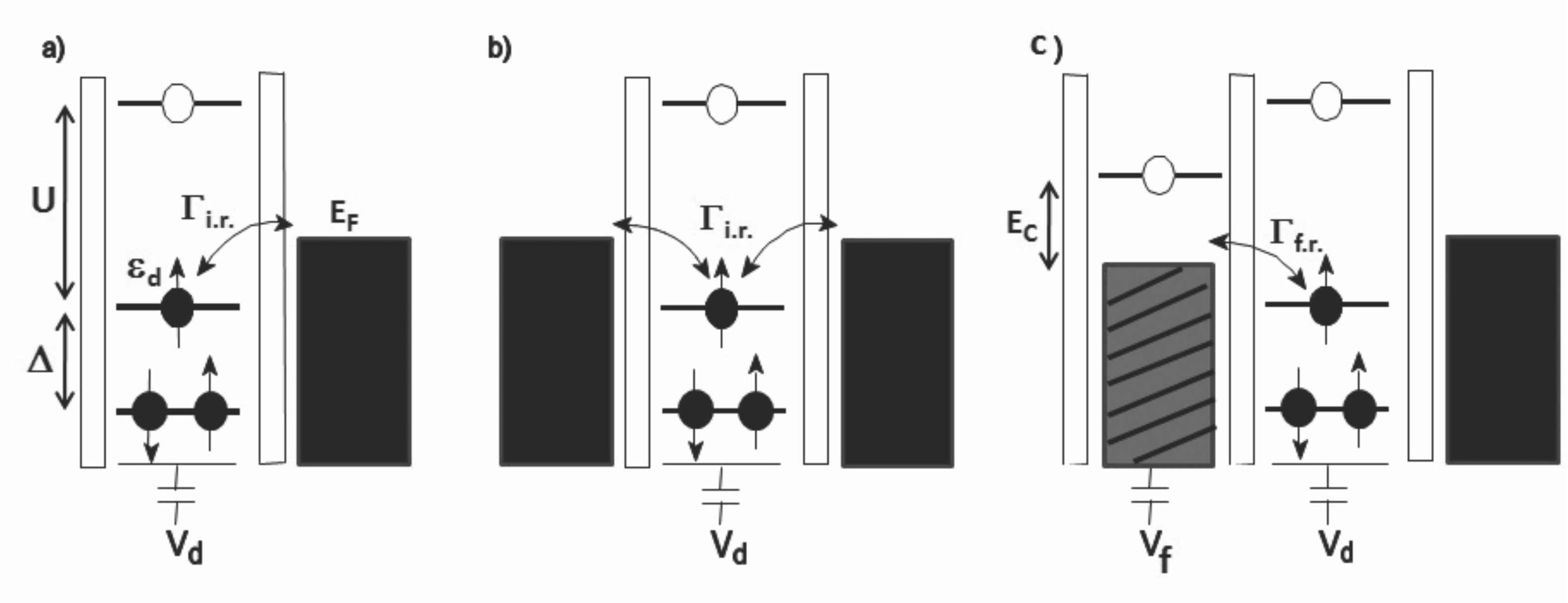}
}\end{center}
\caption{(a) Tunneling $\Gamma_{i.r.}$ between a quantum dot and a nearby reservoir gives rise to
antiferromagnetic exchange coupling between the local spin and the spins in
the reservoirs. (b) Physically separated reservoirs are not necessarily independent: if a localized electron can hop off the site to the right reservoir and a new electron can hop onto the site from the left, the two reservoirs will cooperate in screening the localized spin. (c) Adding a second reservoir that
also antiferromagnetically couples to the local spin: Coulomb
charging on the second (finite) reservoir causes the two reservoirs, labeled $i.r.$ (infinite) and $f.r.$ (finite), to screen the local
spin independently. If the couplings of the two reservoirs are equal, this results in the overscreening characteristic of the two-channel Kondo effect.}
\label{fig:Jterms}
\end{figure}

The first experimental observation of the 2CK state in a quantum dot system  occurred in the double dot geometry (Fig.~\ref{fig:Jterms} (c)) proposed by Oreg and Goldhaber-Gordon~\cite{dgg.2ck}, though there were earlier related proposals (see Sec.~\ref{sec:TwoCKOtherGeo}). The Hamiltonian that describes this system is that of Eq.~(\ref{eq:modelhamiltonian}). The localized magnetic impurity is represented by a small quantum dot containing an odd number of electrons. The conduction electrons that screen this local moment belong to two reservoirs, as illustrated in Fig. \ref{fig:Jterms}. One of the reservoirs corresponds to the source and drain leads (``left" and ``right" electrons), which, although physically separated, form a single effective reservoir ~\cite{glazmanraikh.kondo}, denoted by $i.r.$ for infinite reservoir. The second reservoir, denoted by $f.r.$ is a finite electron bath, ie., a much larger quantum dot with fixed electron occupancy. It constitutes a second independent screening channel (striped in Fig.~\ref{fig:Jterms} (c)) at low temperature when Coulomb Blockade forbids it from exchanging electrons with the infinite reservoir (black in Fig.~\ref{fig:Jterms} (c)).

The QPT of Fig.~\ref{fig:Jterms} occurs as a function of the relative coupling of the two channels to the localized impurity. The antiferromagnetic coupling between the impurity and each reservoir depends on the tunneling rate to that reservoir and the depth of the impurity level with respect to that reservoir's Fermi level; this coupling determines the reservoirs' respective ability to Kondo-screen the local moment. Two processes, one involving electrons and one involving holes, contribute to Kondo screening. In energies $E^n_l$ to follow,  the subscript refers to the number of electrons on the dot and the superscript refers to the change in the number of electrons in the finite reservoir. To first order, the virtual exchange of electrons with the local moment requires an energy $\Delta E_{e, i.r.}=E^{0}_2- E_1^0$ for spin-flip events with the infinite reservoir, and an energy  $\Delta E_{e, f.r.}=E^{-1}_2- E_1^0$  for spin-flips with the finite reservoir. The equivalent hole process between the dot and the infinite reservoir requires an energy difference $\Delta E_{h,  i.r.}=E^0_0- E_1^0$ or  alternatively, an energy $\Delta E_{h,  f.r.}=E^1_0- E_1^0$ for spin-flip events between the dot and the finite reservoir.  Therefore, the two exchange couplings are given by \cite{dgg.2ck}
\begin{equation} \label{yuvaleq1}
J_{i.r.} = \Gamma_{i.r.} \left(\frac{1}{E_2^0 - E_1^0} +
\frac{1}{E_0^0 - E_1^0}\right)
\end{equation}
\begin{equation} \label{yuvaleq2}
J_{f.r.} = \Gamma_{f.r.} \left(\frac{1}{E_2^{-1} - E_1^0} +\frac{1}{E^1_0 - E_1^0} \right)
\end{equation}

At zero temperature the system is in one of the three distinct ground states:  a Fermi liquid (Kondo singlet state with the finite reservoir) when $J_{f.r.}>J_{i.r.}$, another Fermi liquid (Kondo singlet state with the infinite reservoir) when $J_{i.r.} > J_{f.r.}$, and the nFL 2CK state at the QCP ($J_{i.r.} =  J_{f.r.}$).

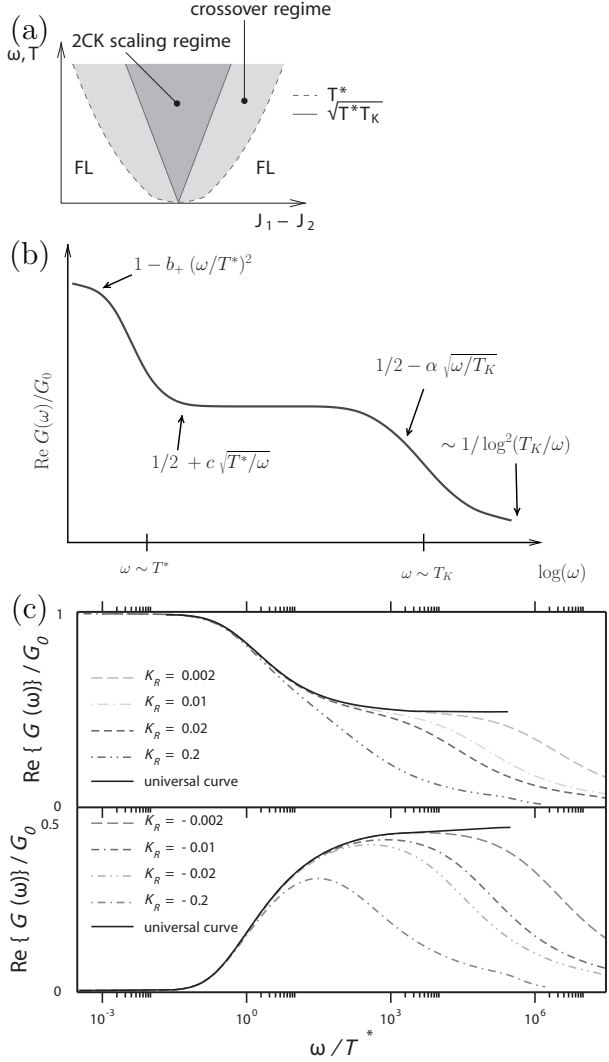
\begin{figure}
\setlength{\unitlength}{1cm}
\begin{center}
\begin{picture}(12,14)(0,0)
%\put(0,0){\line(0,1){8.6}}
\put(0,11){\includegraphics[width=5.0cm,keepaspectratio=true]{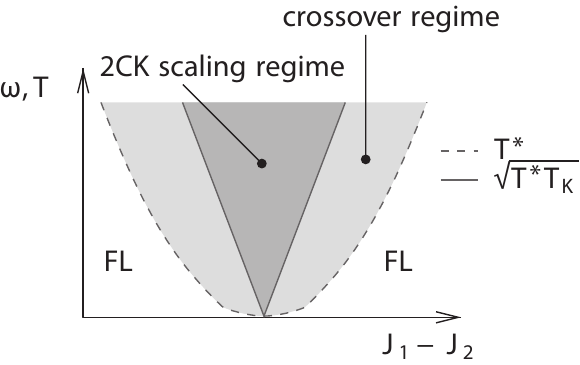}}
\put(0.3,13.7){\makebox(0,0){(a)}}
\put(0,6.1){\includegraphics[width=8cm,keepaspectratio=true]{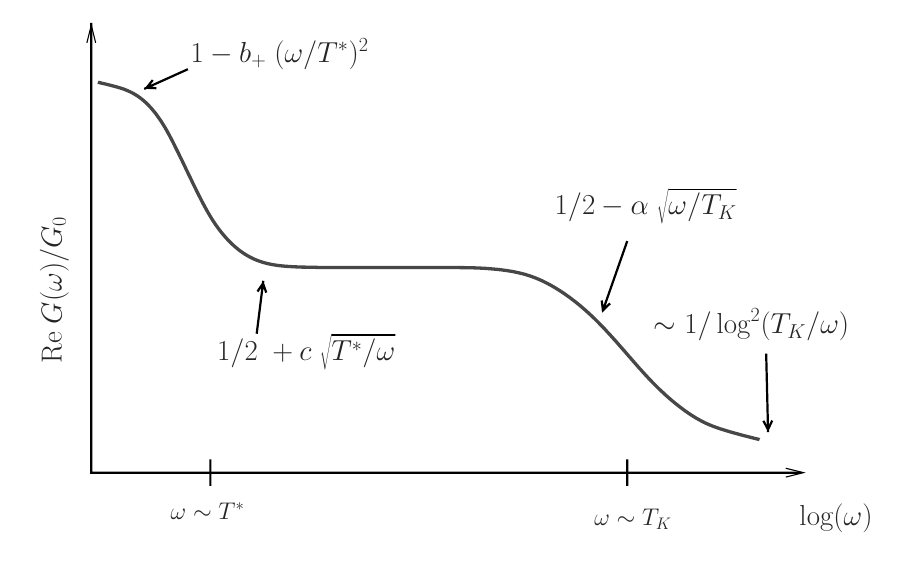}}
\put(0.3,10.7){\makebox(0,0){(b)}}
\put(0,0){\includegraphics[width=8cm,keepaspectratio=true]{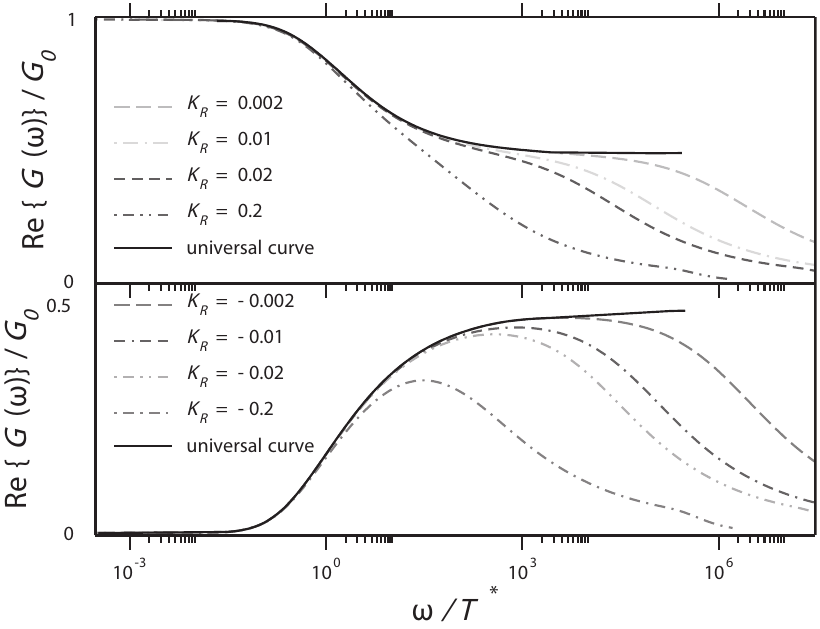}}
\put(0.3,6){\makebox(0,0){(c)}}
\end{picture}
\end{center}
\caption{(a) A more detailed look at the phase transition: a cross-over region separates the 2CK scaling region from the 1CK scaling region. (b) Sketch of the AC conductance as a function of frequency for $\Delta J > 0$ . The stated dependences in the different regions are based on conformal field theory~\cite{zarand.2ck}. (c) Numerical renormalization group calculations of the real part of the conductance in the universal cross-over regime. Reprinted figures with permission from [T\'{o}th \textit{et al.} Phys. Rev. B 76, 155318 (2007)]. Copyright (2007) by the American Physical Society.
}
\label{fig:Crossover}
\end{figure}

At finite temperature, the presence of the QCP manifests itself in the scaling behavior of the system. For temperatures below the Kondo temperature but above the energy scale  $T_\Delta$ associated with the channel anisotropy $\Delta J = J_{i.r.} - J_{f.r.}$ (see Sec.~\ref{sec:AnisotropyBfield} )
the system will obey the 2CK scaling form of Eq.~(\ref{eq:ScalingCurves2}). This is the quantum critical region. On the two sides of the quantum critical region, where  $T< T_\Delta$, the scaling law for single channel Kondo given by Eq.~(\ref{eq:ScalingCurves}) is expected. When the infinite reservoir is better coupled, the Kondo effect manifests as an enhancement around zero bias (and zero temperature), compared to the conductance at higher bias and temperature. Conversely, when the finite channel is better coupled to the impurity, it effectively steals the Kondo state away from the leads. This results in a suppression rather than enhancement of the conductance through the small quantum dot at low bias and temperature. Recent work suggests that a crossover region divides the 1CK Fermi liquid from the 2CK nFL regime \cite{zarand.2ck}. This is the lightly shaded region in Fig.~\ref{fig:Crossover}  and is discussed in more detail in Sec.~\ref{sec:AnisotropyBfield}.

\begin{figure}[t] \begin{center}
\resizebox{9cm}{!}{
\includegraphics{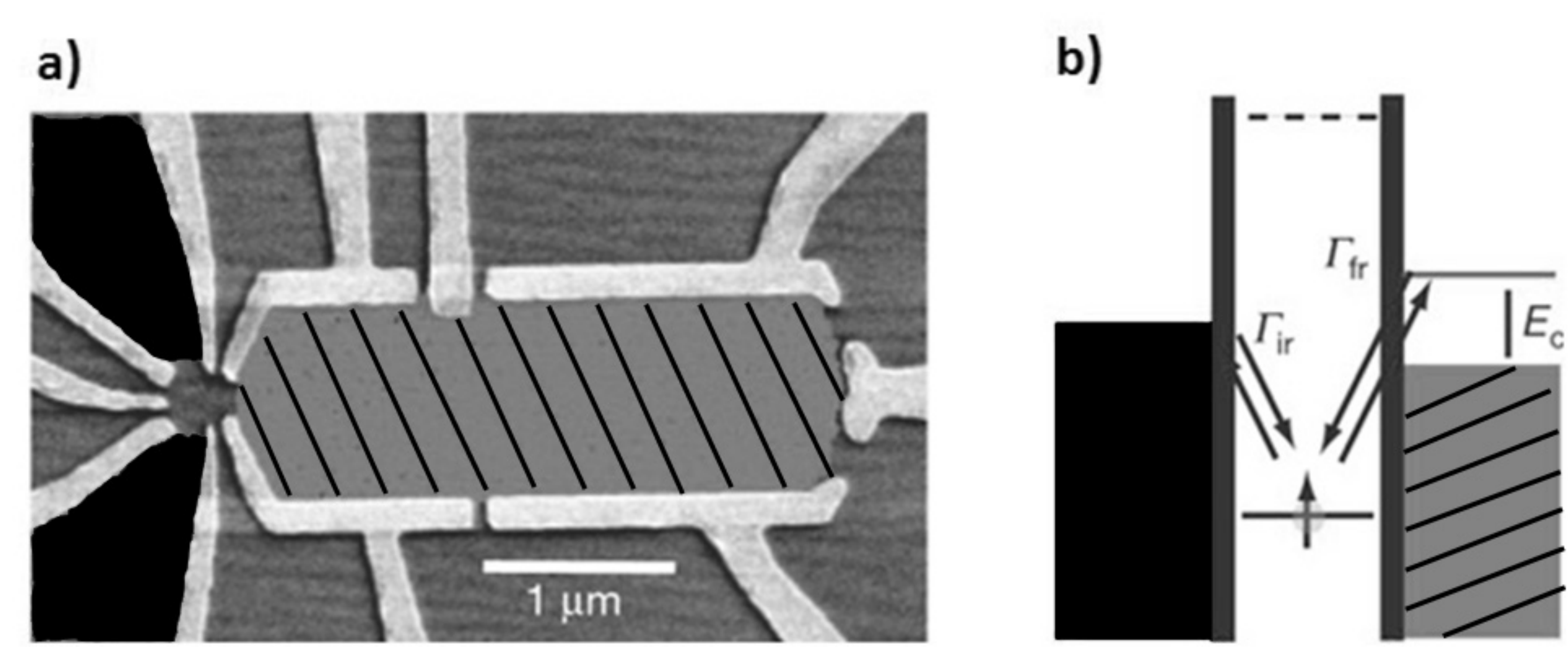}
}\end{center}

\caption{(a) Annotated SEM image of the two-channel Kondo nanostructure: an additional finite reservoir (striped) is coupled to an artificial magnetic impurity already connected to an infinite reservoir composed of two conventional leads (black). (b) Coulomb blockade suppresses exchange of electrons between the finite reservoir and the normal leads at low temperature. The two reservoirs hence act as two independent screening channels. Figure adapted from Potok \textit{et al.}~\cite{ron}.}
\label{fig:device}
\end{figure}

This 2CK system was realized in a lateral GaAs quantum dot geometry (Fig.~\ref{fig:device} (a)). The  metal gates patterned on the surface of the GaAs/AlGaAs heterostructure define the small quantum dot and the finite reservoir, and allow precise tuning of both the electrostatic potential of each quantum dot and the tunneling rates to the two reservoirs.

The small quantum dot contains $\approx 25$ electrons and acts as the Kondo impurity. Its bare charging energy is $\approx$1
meV (corresponding to $\approx$10 K) and the single-particle energy level spacing is $100~\mu$eV. It is coupled to a set of leads (black regions in Fig.~\ref{fig:device}(a) and (b)) that form a bound singlet state with the dot, with $k T_{K}\approx 13~\mu$eV. The second independent channel consists of electrons in the $3~\mu m^2 $ dot (striped region in Fig.~\ref{fig:device}(a) and (b)) which corresponds to a charging energy $E_c$ of $100~\mu eV$. This prevents the exchange of electrons with the other reservoir when the large dot is tuned to a Coulomb Blockade valley (a well-defined charge state) since the lowest electron temperature is  $T_{\rm base} = 12~$mK. Yet the large dot effectively has a continuum of single-particle states: its single-particle energy level spacing of $\approx 2~\mu$eV cannot be resolved at base temperature since $kT_{\rm base}\approx1~\mu$eV. Thus the finite reservoir acts as a second, independent Kondo-screening channel, as required for observing the 2CK effect.

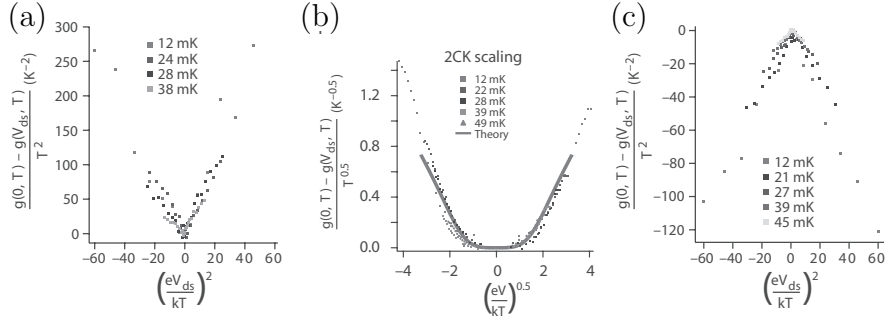
\begin{figure}[t]
\setlength{\unitlength}{1cm}
\begin{center}
\begin{picture}(12,4.4)(0,0)
%\put(0,0){\line(0,1){4.3}}
\put(0,0){\includegraphics[width=4cm,keepaspectratio=true]{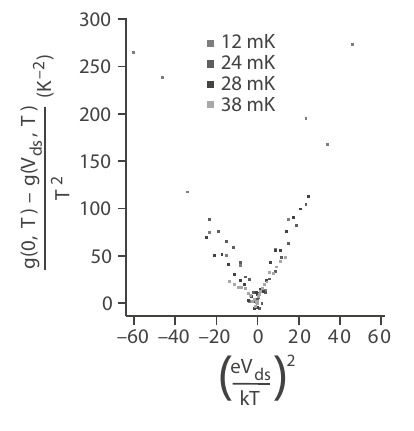}}
\put(0.4,4.1){\makebox(0,0){(a)}}

\put(4,0){\includegraphics[width=4cm,keepaspectratio=true]{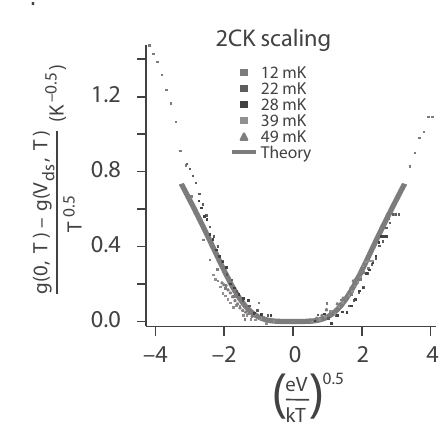}}
\put(4.3,4.1){\makebox(0,0){(b)}}

\put(8,0){\includegraphics[width=4cm,keepaspectratio=true]{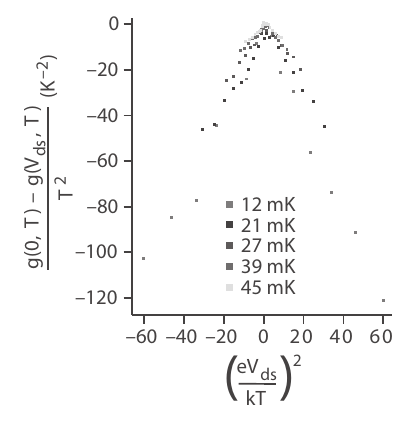}}
\put(8.4,4.1){\makebox(0,0){(c)}}
\end{picture}
\end{center}

\caption{(a) Measurements showing 1CK Kondo scaling of the data for $J_{i.r.}>J_{f.r.}$ (b) Measurements showing 2CK scaling of the data when $J_{i.r.}=J_{f.r.}$ (c) Measurements showing 1CK scaling in the $J_{i.r.}<J_{f.r.}$ case. Figure adapted from Potok \textit{et al.}~\cite{ron}.
}
\label{fig:scaling}
\end{figure}

The measurement of the conductance through the quantum dot for three different couplings to the two channels reveals the different scaling regions of the phase diagram in Fig.~\ref{fig:2CKwithPetals}(a). Voltages on the different gates in Fig.~\ref{fig:device} (a) are used to control the small and large dot energy levels relative to the Fermi levels in the leads as well as the overall tunnel coupling; together these determine the exchange coupling to the finite and infinite reservoirs(Eq.~(\ref{yuvaleq1}) and (\ref{yuvaleq2})).
When the infinite reservoir is much more strongly coupled to the dot than is the finite reservoir, the zero bias conductance enhancement characteristic of 1CK is observed (Fig.~\ref{fig:scaling} (a)).  Conversely, for increased interdot tunneling ($J_{i.r.}<J_{f.r.}$), the 1CK state is formed with the finite reservoir  and causes the suppression of the conductance in Fig.~\ref{fig:scaling} (c). In both cases, the differential conductance at different temperatures and biases scales according to Eq.~(\ref{eq:ScalingCurves}) as shown in Fig.~\ref{fig:scaling} (a) and (c). The exponent $\alpha $ from the scaled conductance $(G(0, T) - G(V_{ds}, T))/T^{\alpha}$ is numerically determined to be 1.72 $\pm $ 0.40, which is consistent with the value $\alpha = 2$ characteristic of the 1CK Fermi liquid ground state and the exponent in Eq.~(\ref{eq:ScalingCurves}).

The evolution of the differential conductance  with tuning from the 1CK state with the infinite reservoir to the 1CK state with the finite reservoir doesn't show clear $\sqrt{V_{ds}}$ behavior as one would expect when crossing through the 2CK point. However, Fig.~\ref{fig:scaling}(b) shows that the scaled conductances $(G(0, T) - G(V_{ds}, T))/T^{\alpha}$ at different temperatures collapse onto the same line when $\alpha=0.5$, as expected for the 2CK nFL state and in agreement with Eq.\ref{eq:ScalingCurves2}. Single-channel Kondo scaling of the same data fails \cite{ron}. A two dimensional nonlinear fit to the data in Fig.~\ref{fig:scaling} (b) produced the value $\alpha = 0.62 \pm 0.21$, consistent with 2CK
behavior.

\subsection{The Two-channel Kondo Effect in other Quantum Dot Geometries}
\label{sec:TwoCKOtherGeo}
For the three distinct Kondo states involved in the QPT described above, the degeneracy is provided by the unpaired spin-$\frac{1}{2}$ in the small quantum dot. There are several theoretical proposals for the realization of the 2CK effect where the local degeneracy is based on the charge degree of freedom. In fact, the earliest proposal for observing the 2CK effect in semiconductor nanostructures~\cite{matveev.other1ck} involved a large semiconductor quantum dot coupled via single-mode point contacts to a reservoir. Here, at the charge degeneracy points of the dot, strong charge fluctuations are expected to give rise to a 2CK effect~\cite{matveev.1ck, lebanon.2ck}: two successive charge states play the role of the local two-fold degeneracy, while the two independent screening channels are the spin-up and spin-down electrons of the reservoir. Due to conflicting constraints on the size of the dot this specific proposal may not be experimentally realizable \cite{zarand.matv2ck}.

This difficulty can be overcome by introducing a single resonant level between the large quantum dot and the reservoir. At the charge degeneracy point of the large dot, and in the mixed-valence regime of the small dot, a 2CK effect with a nFL fixed point is expected to occur ~\cite{Lebanon.other2ck}. This modified geometry is in fact identical to the double dot geometry described above, where spin-2CK is observed away from the charge degeneracy points of both dots. In fact, further analysis of the different parameter regimes of this double dot system predicts several exotic effects such as a line of two-channel fixed points, a continuous transition from the spin-2CK to the charge-2CK effect~\cite{anders.2ck} and an SU(4) Kondo effect with a stable fixed point~\cite{LeHur.other2ck}.

\subsection{The Two-channel Kondo Effect in Graphene Sheets}
In pristine graphene, there are always two angular momentum channels that couple to a local moment, so when the exchange coupling is the same for the two channels 2CK behavior is expected. Recently, researchers studying tunneling into Cobalt adatoms on graphene observed a conductance scaling law consistent with that expected for 2CK physics \cite{Mattos2009}. In this system the Fermi level was far from the Dirac point (in Sec \ref{sec:Reservoirs} we will discuss the effects of suppressing the density of states near the Fermi energy).

Low-temperature scanning tunneling spectroscopy measurements on epitaxial
graphene monolayers exhibiting a linear Dirac dispersion, dosed with a low coverage of Cobalt
atoms, revealed two kinds of Kondo resonances. The difference between
them is associated with whether the Co atom occupies the middle of a
hexagon in the graphene lattice or is located on top of a carbon atom.
The spin of the Cobalt atom is calculated \cite{Wehling2009_CoOnGraphene, Yagi2004_dopedCNT} to be 1/2. The degenerate K and K'
valleys in graphene result in two screening channels for the localized
magnetic impurity. When the adatom is in the center of a hexagon, it
should be equally coupled to the K and K' valleys, resulting in
2CK behavior. In this case, scaling of differential
conductance with tip bias at constant low temperature (4 K) over a
bias range from $V=kT$ to $V=kT_K$ was observed to be consistent with the
exponent 0.5 expected for 2CK. In contrast, when the Co
atom is located on top of a carbon atom it should locally break the
valley symmetry and thus show single channel Kondo physics. The
exponent extracted from the voltage scaling in this case is close to the Fermi
liquid value of 2. This also agrees with measurements of Cobalt atoms
on Copper which show single channel Kondo physics \cite{Mattos2009}.

\subsection{The Two-impurity Kondo Effect in a Double Quantum Dot Geometry}
\label{sect:TwoIK}

The single-channel and two-channel Kondo effects describe the behavior of a single magnetic impurity interacting with conduction electrons. In bulk materials with magnetic impurities, this corresponds to the dilute limit, where the density of magnetic impurities is sufficiently low that the local spins do not interact with each other. At the opposite extreme there are materials, such as Kondo lattice materials based on Cerium and Uranium, where local moments are closely spaced~\cite{hewson}. In these materials, interactions between the impurities are mediated by the conduction electrons via the Ruderman-Kittel-Kasuya-Yosida (RKKY) mechanism. The unique properties of these materials arise from the competition between the Kondo screening of individual impurities and the inter-impurity interactions, and studying this competition is important to gaining a more complete understanding of these materials.

As a first step toward tackling this complex problem, consider a simpler system which can be realized in a nanostructure: just two magnetic impurities coupled to conduction electrons and to one another, known as the 2-impurity Kondo system (2IK) \cite{Jones1987:2MagImp}. The model Hamiltonian for this system is
\begin{equation}
H_{\rm 2IK}= J_1 s_1 \cdot S_1 + J_2 s_2\cdot S_2 + K S_1 \cdot S_2
\end{equation}
where $S_1$ and $S_2$ are the spins of the two magnetic impurities, $s_1$ and $s_2$ are the spins of the conduction electrons at the impurity sites, and $J_1, J_2 >0$ are the anti-ferromagnetic (AFM) interaction between the conduction electrons and the impurities that cause the Kondo screening of the impurities. $K$ is the inter-impurity interaction, from the RKKY mechanism. We consider the case where $T_{K,1} \sim T_{K,2} \sim$ a single effective $T_{K}$ for the full system. For $-T_K< K < K_c$ (where $K_c\approx 2.5 T_K$) the Kondo effect dominates and each impurity is individually screened by the conduction electrons. However, for $K> K_c$ the inter-impurity interaction dominates and the two impurities pair to form a spin singlet that does not interact with the conduction electrons\cite{Jones1987:2MagImp}. At the critical coupling $K= K_c$ between the Kondo-screened and singlet ground states the system is predicted to go through a QPT associated with a novel nFL fixed point \cite{Jones1989:MF2IK, Jones1989:CritPt, Affleck1992:Exact2IK}.

Nanostructures offer a natural means of realizing the 2IK Hamiltonian (see \cite{Chang2009_KondoInCoupledDots} and references therein). The 2IK Hamiltonian has been modeled experimentally with gated semiconductor quantum dots: Craig \textit{et al.} \cite{craig.2ikm} have studied two quantum dots coupled by an intermediate reservoir of electrons, while Jeong \textit{et al.} \cite{twoimpkondo.exp} and Chen \textit{et al.} \cite{Chen2004:QStateTransition} have studied a double quantum dot system with no intervening reservoir, where the AFM coupling is provided by the exchange interaction between the two dots. These authors found a suppression of the Kondo conductance of a dot as they increased the strength of the AFM inter-dot coupling, evidence that the formation of an inter-impurity spin singlet was disrupting the Kondo coupling to the leads and hence conduction through the dots. Heersche \textit{et al.} \cite{heersche.2ikm} studied transport through a gold grain quantum dot in the presence of cobalt impurities which could couple to a spin on the grain. They observed a similar suppression of the Kondo conductance in the presence of the cobalt impurities, and found that the Kondo peak could be restored by application of a magnetic field. These observations show that one can tune a 2IK system to either side of the QPT. However, the nFL behavior at the QCP has yet to be observed.

\subsection{The Two-impurity Kondo Effect in a Quantum Dot at the Singlet-Triplet Transition}

The 2IK Hamiltonian can be also used to describe a quantum dot at the singlet-triplet degeneracy point~\cite{kikoin.2ik}. Therefore, a QPT of the Kosterlitz-Thouless type should also take place in a single dot geometry, where the quantum dot has an even number of electrons and is coupled to one electron reservoir.

The ground state of a two-electron-system is a spin singlet ($S=0$), with both electrons occupying the same orbital.  The triplet state ($S=1$) will have lower energy if the exchange energy gained for parallel spin filling exceeds the level separation between adjacent orbitals ~\cite{tarucha.00}. %The case of a quantum dot with S=1 is different from the case of a spin-$\frac{1}{2}$ because the screening electrons no longer have the same spin symmetry as the local site. I
Instead of the fully screened or overscreened scenarios of the 1CK or 2CK cases, for a single reservoir mode coupled to the S=1 quantum dot, Kondo correlations only partially screen the local impurity and a residual spin-$\frac{1}{2}$ is left over. Therefore, an underscreened Kondo effect takes place on the triplet side.

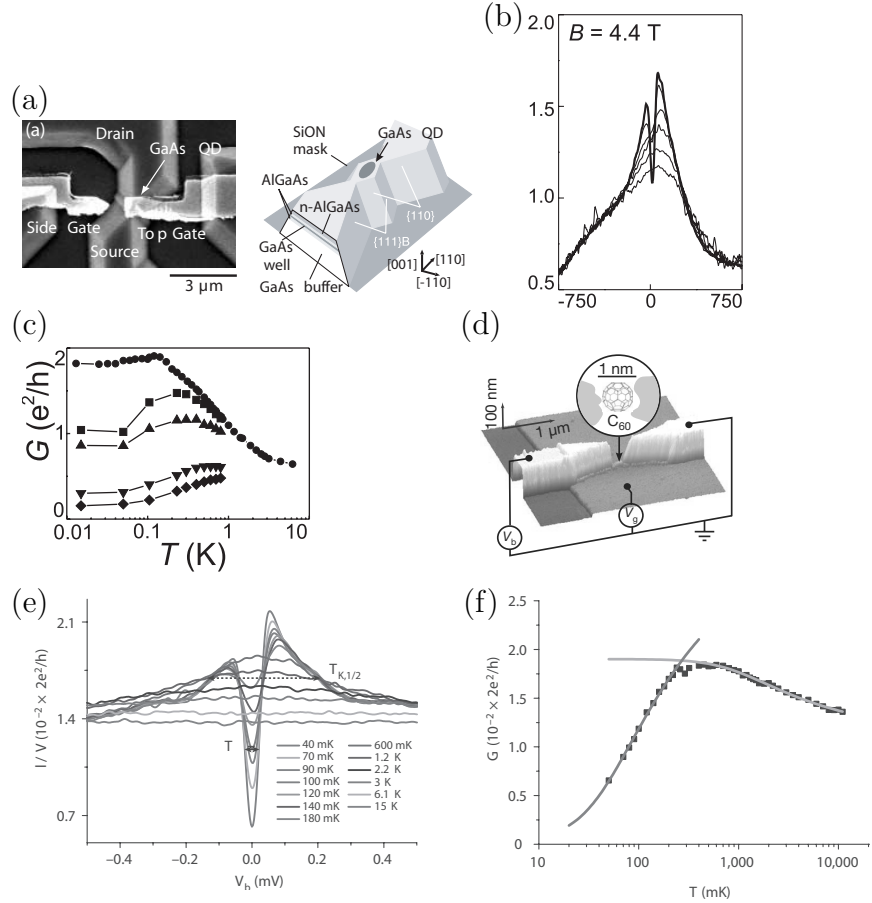
\begin{figure}
\setlength{\unitlength}{1cm}
\begin{center}
\begin{picture}(12,12.1)(0,0)
\put(0,8.2){\includegraphics[width=6.0cm,keepaspectratio=true]{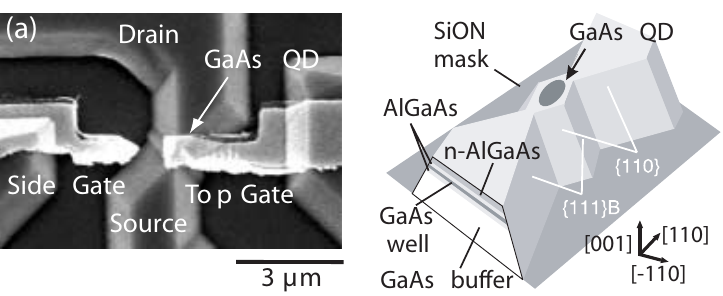}}
\put(0.1,10.8){\makebox(0,0){(a)}}

\put(6.6,7.8){\includegraphics[width=3.0cm,keepaspectratio=true]{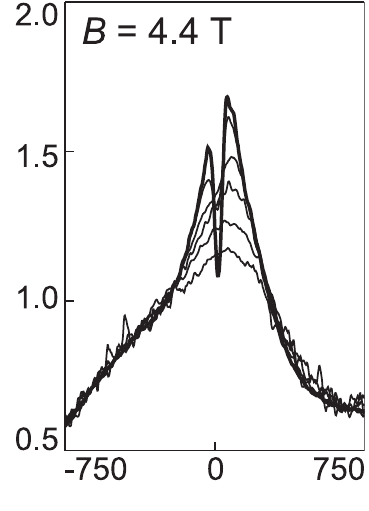}}
\put(6.4,11.9){\makebox(0,0){(b)}}

\put(0,4.5){\includegraphics[width=4cm,keepaspectratio=true]{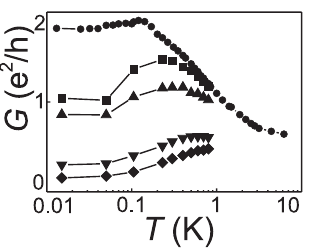}}
\put(0.1,7.8){\makebox(0,0){(c)}}

\put(6,4.7){\includegraphics[width=4cm,keepaspectratio=true]{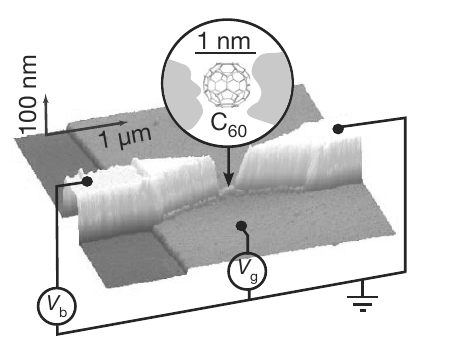}}
\put(6.1,7.8){\makebox(0,0){(d)}}

\put(0,0.3){\includegraphics[width=5.5cm,keepaspectratio=true]{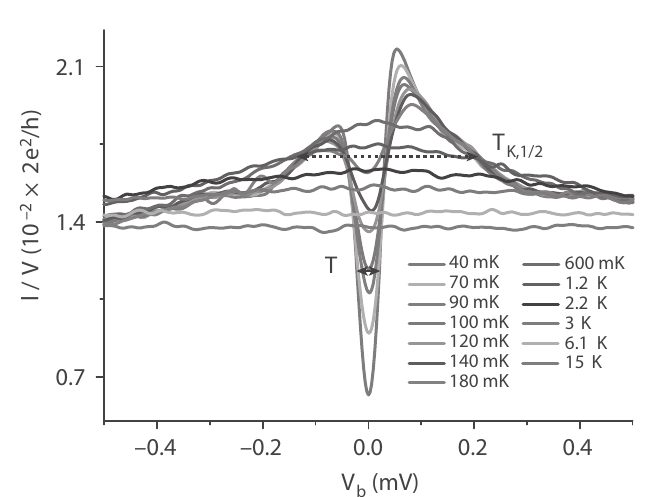}}
\put(0.1,4.1){\makebox(0,0){(e)}}

\put(6,0){\includegraphics[width=5.5cm,keepaspectratio=true]{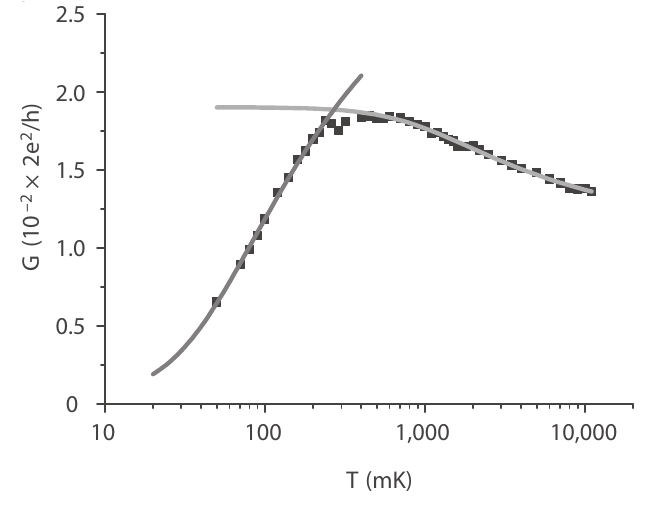}}
\put(6.1,4.1){\makebox(0,0){(f)}}

\end{picture}
\end{center}
\caption{(a) SEM and schematic of a lateral quantum dot device. (b) Conductance on one side of the degeneracy point as a function of bias for different temperatures. The zero-bias anomaly develops a dip with decreasing temperatures. (c) The logarithmic dependence of the conductance on temperature as the device is tuned away from the singlet-triplet transition point.
Reprinted figures with permission from [W. G. van der Wiel, \textit{et al.} Phys. Rev. Lett. 88, 126803 (2002)]. Copyright (2002) by the American Physical Society. (d) Sketch of C-60 single molecule transistor. On the singlet side, the characteristic dip in the zero-bias conductance enhancement (e) as well as the logarithmic enhancement and suppression of the conductance as a function of temperature (f) is observed. Adapted by permission from Macmillan Publishers Ltd: Roch \textit{et al.}, Nature 453, 633 (2008).
}
\label{fig:vdWAndR}
\end{figure}

The energy difference between the singlet and the triplet states can be controlled by applying a magnetic field or by tuning the gate voltage. A singlet-triplet Kondo effect has been observed in transport measurements through vertical quantum dots, lateral quantum dots, carbon nanotubes, and single molecule transistors (see \cite{roch.st}, \cite{charis.kondo}, \cite{ReviewKondo} and references therein).

A system near a singlet-triplet transition can exhibit a two-stage Kondo effect~\cite{phase.hof,pustilnik.st03} which is characterized by a logarithmic and non-monotonic dependence of the conductance on temperature and bias. When two reservoir channels couple to the spin-1 impurity, two Kondo-screening channels with two different Kondo temperatures $T_{K1}$ and $T_{K2}$ form \cite{pustilnik.st01}.  With decreasing temperatures, the conductance is enhanced when $T_{K2}<T<T_{K1}$ as half of the spin 1 is screened by one channel. Once  $T < T_{K2}<T_{K1}$ the residual spin-$\frac{1}{2}$ is screened by the second channel, and the conductance is reduced by the interference of the two modes. On the singlet side of the degeneracy a similar two stage Kondo effect is predicted to occur if only one channel is coupled to the quantum dot \cite{phase.hof}.

The transition from the singlet to the triplet state of a quantum dot is very different, depending on which Kondo state is present on each side. When one reservoir is coupled, the transition happens from a two-stage Kondo state on the singlet side to an underscreened Kondo state on the triplet side via a QCP of the same type as that of the 2IK model ~\cite{phase.hof}. When two reservoirs couple to the quantum dot, the transition from a spin zero system on the singlet side to a two-stage Kondo system on the triplet side is a crossover, and the ground state has Fermi liquid properties throughout ~\cite{pustilnik.st00, pustilnik.st03}.

This effect has  been observed in lateral quantum dots~\cite{twostage,granger.st} and single molecule transistors~\cite{roch.st}.
In the experiment by van der Wiel \textit{et al.}~\cite{twostage} a lateral quantum dot is tuned near a singlet-triplet degeneracy point using magnetic field. Conductance measurements show a dip in the zero-bias anomaly at low temperatures, which disappears with increasing temperature (Fig.~\ref{fig:vdWAndR}(b)), as well as the predicted logarithmic increase and decrease of the conductance with temperature and or bias (Fig.~\ref{fig:vdWAndR}(c)). However, in this system it is not possible to unambiguously distinguish whether a singlet or triplet ground state is responsible for the observed effect. This makes it impossible to determine which of the two proposed mechanisms is responsible.

Roch \textit{et al.} \cite{roch.st} have measured transport through a C-60 single electron transistor shown in Fig.~\ref{fig:vdWAndR}(d), and used a gate voltage to tune through the singlet-triplet transition. Using bias spectroscopy as a function of both gate voltage and magnetic field, they are able to clearly identify the singlet and triplet sides of the transition. In Fig~\ref{fig:vdWAndR}(e) they observe a zero-bias dip in the conductance on the singlet side of the transition, whose width narrows as the gate voltage is tuned toward the transition. The zero-bias conductance of the dip increases logarithmically with increasing temperature (Fig.~\ref{fig:vdWAndR} (f)), and conductance dips at different distances from the transition can be scaled to lie on a single curve. On the triplet side, they observe a zero-bias peak in conductance, which also has a logarithmic temperature dependence. This behavior, as well as the scaling of the conductance with temperature and bias, is consistent with predictions for a QPT between the singlet and the triplet regions.

\section{Looking Forward}

  Having reviewed the progress in observing QPTs in nanostructures, we now look forward and examine some of the most promising proposals for characterizing existing QPTs and finding new QPTs in novel and as yet unrealized nanostructures.

\subsection{Influence of  Channel Asymmetry and Magnetic Field on the Two-channel Kondo Effect}
\label{sec:AnisotropyBfield}
The experiments in Sec. \ref{sec:TwoCK_DQD} demonstrated that one can tune to the region where the asymmetry parameter $\Delta J= J_{i.r.} - J_{f.r.}$ is large and 1CK behavior dominates, and to the point where $\Delta J\approx 0$ and one observes 2CK behavior. The next important step is to map out the full phase diagram sketched in Fig. \ref{fig:Crossover}(a). Of especial interest is the region around $\Delta J= 0$ where the asymmetry is small but finite: here theory predicts that the proximity of the QCP influences the behavior of the system at both low and high energy \cite{zarand.2ck}. Specifically, a new energy scale $T_\Delta \propto (\Delta J)^2$ emerges. In the region $T>(T_\Delta T_K)^{1/2}$ (dark gray in Fig. \ref{fig:Crossover}(a)) the 2CK scaling law should hold. However, for $T_\Delta < T < (T_\Delta T_K)^{1/2}$ it is predicted that one should enter a cross-over region (light gray in Fig. \ref{fig:Crossover}(a)) where conductance is described by a new universal function that has been calculated numerically (Fig. \ref{fig:Crossover}(c)). Finally, for $T<T_\Delta$ it is predicted that one should observe a modified 1CK scaling law, where the relevant scaling parameter is $T/T_\Delta$, not $T/T_K$. The full evolution is sketched in Fig. \ref{fig:Crossover}(b). This phase diagram can also be accessed with bias replacing temperature, and together these experimental knobs should allow a complete mapping of the phase diagram.

The QCP can also be accessed at $\Delta J= 0$ as a function of magnetic field: unlike for 1CK, Zeeman splitting is a relevant perturbation and any non-zero magnetic field will move the system away from the QCP. The phase diagram is similar to the one shown in Fig.~\ref{fig:2CKwithPetals}(a) for a transition driven by channel anisotropy, but here Fermi liquid behavior should appear below an energy scale given by $E_{B}= (g \mu B)^2/k T_{k}$, and the functional form of the conductance curves will be different from the anisotropy-driven case. The contrast between 1CK and 2CK behavior is manifested in the effect of the field on the Kondo peak in the spin-averaged density of states (differential conductance) as a function of bias. For 1CK, there is a threshold magnetic field $B_{\mathrm{thresh}}$ above which the Kondo peak will split \cite{ReviewKondo}, but for $B<B_{\mathrm{thresh}}$ no splitting is observed. In contrast, for 2CK it is predicted \cite{zarand.2ck} that the Kondo peak remains split even down to infinitesimal fields. This measurement is experimentally challenging because the magnetic field needs to be applied parallel to the two-dimensional electron gas at the GaAs/AlGaAs interface. If there is a sizable component of the field perpendicular to the interface, then this component will couple to the orbital motion of the electrons and modify the wavefunction overlap of the dot and its leads. The result of this will be to disrupt the delicate tuning necessary to maintain $\Delta J= 0$, making it impossible to measure the desired splitting. Such field alignment is well established but non-trivial.

\subsection{Multiple Sites}
A natural extension beyond two neighboring local sites is three sites
arranged as vertices of an equilateral triangle.  Such a system can be
built either by using STM to position atoms or molecules on a surface
(e.g. Cr$_3$ on Au(111)~\cite{jamneala}, Co$_3$ on
Cu(111)~\cite{manoharan.communication}, or a linear Mn chain on CuN on Cu(111)~\cite{hirjibehedin} or using laterally-coupled quantum dots, see
e.g.~\cite{vidan, gaudreau, rogge, amaha, schroer}. But even after defining the geometry there are many
options for the Hamiltonian and its parameters: occupancy and spin of
each site, strength of antiferromagnetic coupling between pairs of
sites and between a site and mobile conduction electrons, symmetry of
local site energies and couplings, etc.

This has given rise to a wide
range of theoretical proposals, many of which are reviewed in~\cite{kikoin.triple}. Without trying to survey all the
theoretical contributions, we point out a few that are particularly
intriguing to us. In 2005, Ingersent {\em et al.} and Lazarovits {\em et
al.} considered three antiferromagnetically-coupled half-integer spins
arranged in a triangle, and coupled to conduction electrons by a
standard Kondo coupling~\cite{ingersent.triple,
lazarovits}. They found that the frustration of the antiferromagnetic
coupling led to a net local spin of 1/2, and a novel non-Fermi-liquid
phase when the Kondo coupling is included. Ingersent {\em et al.}
found that this phase should be stable to asymmetries that would
likely occur in a realistic quantum dot-based realization of this
system, even with careful tuning, whereas Lazarovits {\em et al.}
found that the phase would not be stable to spin-orbit coupling.
Further decreasing the symmetry of couplings between sites and to
leads, Mitchell {\em et al.}~\cite{mitchell.triple} found a QPT between an
antiferromagnetically Kondo-coupled ground state and a
ferromagnetically-coupled (local moment) Kondo ground state, as a
function of the coupling between two particular dots. Restoring the
symmetry of the trimer but increasing the tunnel couplings to a
"molecular regime", Vernek {\em et al.} found a Fermi liquid S=1 Kondo
ground state~\cite{Vernek.triple}. Going beyond three
impurities, and beyond transport spectroscopy (the main probe
envisioned in the works noted above), Karzig {\em et al.} recently
investigated full counting statistics of transport through a chain of
quantum dots, finding a type of phase transition as a function of
tunnel coupling between dots~\cite{karzig.chain}.

Experimentally, all these systems are at an early stage. An
equilaterally-arranged Cr$_3$ on Au(111) shows no Kondo effect at 7K,
whereas a similar trimer Co$_3$ on Cu(111) shows a Kondo resonance
with an unusual ring-shaped spatial arrangement. The Cr trimer may
have a lower Kondo temperature due to nFL physics, but this has yet to
be verified. With respect to quantum dot trimers, researchers have
mainly determined and tuned the occupancy and coupling of the three
sites, through transport and charge-sensing measurements. The
situation would seem to be ripe for testing the predicted exotic
physics, though this would involve a tour de force measurement: the
number of tunable parameters remains daunting.

\subsection{Different Types of Reservoirs}

\subsubsection{Superconducting Leads and Graphene at the Dirac Point}
\label{sec:Reservoirs}

The ability of the fermions in the reservoir to screen the local moment depends on the density of states of the reservoir around the Fermi energy. In the most generic cases, the effective density of states is smooth and finite around the Fermi energy, and at sufficiently low temperature the conduction electrons will be able to form the spin-singlet state with the impurity spin no matter how small the exchange coupling constant is.

In materials with a non-trivial density of states in the reservoir around the Fermi energy, the Kondo effect can be suppressed: the lack of low energy states prevents screening when the exchange coupling is below some critical coupling. If the density of states is zero over some energy interval around the Fermi energy, as is the case in $s$-wave superconductors, a first order QPT between a Kondo-screened phase and a local moment phase occurs. If the density of states is zero only at the Fermi energy, as is the case for $d$-wave superconductors or single layer graphene at the Dirac point, a second order QPT between the two regimes can take place; however, whether this transition takes place is contingent on the exact power-law dependence of the density of states on the energy relative to the Fermi level \cite{Withoff1990_GaplessPhT, Vojta2006_ImpQPT}.

In particular, when the density of states of the reservoir vanishes linearly with energy, Castanello \textit{et al.} \cite{Cassanello1996_KondoInFluxPhases} predict a QPT between the Kondo-screened phase and the local moment phase, driven by the exchange coupling between the local spin and the conduction electrons. They also predict logarithmic corrections to the Kondo scaling in the Kondo-screened region of the phase diagram. An alternative approach suggests a suppression of this phase transition for certain combination of local degeneracy number and number of channels \cite{GonzalezBuxton1996_GaplessFermiSys}. A physical system that might exhibit such a transition, but which has not yet been realized, is a quantum dot with $d$-wave superconducting leads. This system is of interest because it can help one  to understand the effect of impurities in the cuprates, where even non-magnetic atoms such as Zn may induce magnetic effects \cite{Vojta2002_KondoIndwave, Pan2003_Znimp}.

A quantum dot with $s$-wave superconducting leads has been realized using a nanotube quantum dot contacted with Al leads \cite{Buitelaar2002_SCKondo}, a self-assembled InAs dot with Al leads \cite{buizert2007_SCKondoScaling}, and more recently in a C-60 molecule coupled to Al leads \cite{winkelmann.sc}. In these dots $T_{K}$ could be varied, and the authors observed the expected suppression of the Kondo conductance when $k T_K < \Delta$, where $\Delta$ is the superconducting gap. Surprisingly, for $T_K > \Delta$, it was observed that the Kondo conductance was larger than the normal state conductance \cite{Buitelaar2002_SCKondo}, an effect attributed to the interplay of the Kondo effect and multiple Andreev reflections \cite{Avishai2003_SCQD}.
 The competition between Kondo screening and Cooper pair binding studied in these mesoscopic systems is now being found in real bulk materials: recent work by Sun \textit{et al.} \cite{Sun2009_PressureQPT} suggests that by applying pressure to $\mbox{CeFeAsO}_{\mbox{1-x}}\mbox{F}_{\mbox{x}}$ they could drive a phase transition from a superconducting state where the itinerant Fe $3d$ electrons are in Cooper pairs to a state where these conduction electrons Kondo-screen the local moments of the Ce atoms, thus breaking the superconducting state. Although $\mbox{CeFeAsO}_{\mbox{1-x}}\mbox{F}_{\mbox{x}}$ is not a conventional $s$-wave superconductor, investigating the competition between the Kondo effect and superconductivity in mesoscopic systems offers insight into the fascinating behavior of electrons in real materials.

\subsubsection{The Bose-Fermi Kondo Model in Quantum Dots}
\label{sect:FBK}

In the previous sections, the bath that coupled to the local moment consisted of fermions. There are other systems where the bath consists of bosons, or where a bosonic and a fermionic bath are coupled to the local spin, and the system can undergo a QPT. The case where a spin-1 boson bath and a spin-1/2 fermion bath are coupled to a spin-1/2 local moment is described by the Bose-Fermi Kondo model. When the coupling to the fermionic bath is large, the system undergoes the usual Kondo screening of the local moment by the fermions; when the coupling to the bosonic bath dominates, Kondo processes are suppressed and the systems exhibits universal local moment fluctuations \cite{Smith1999_NFL, Sengupta2000_SpinInFlucField, Zhu2002_CritFluc, Zarand2002_BoseFermiQPT}. Between the two regimes, as a function of the coupling anisotropy to the bosonic/fermionic bath, a continuous phase transition takes place. This model has been proposed to describe some high $T_{\rm c}$ cuprates where the Kondo temperature might be suppressed by the coupling to a bosonic mode, magnetic nanostructures coupled to a metallic bath, or systems near a magnetic QCP \cite{Bobroff1999_SpinlessImpurities, Bobroff2001_LiKondo, Vojta2003_PseudogapFBModel, Millis2001_LocalDefect, Millis2002_QGriffiths}.

In mesoscopic systems, this model can be implemented in a quantum dot coupled to ferromagnetic leads \cite{Kirchner2005_QCinFerro}. Here, a quantum dot with an odd number of electrons plays the usual role of a localized spin-1/2. It is coupled to a bath of conduction electrons, where, due to the nature of the ferromagnetic leads, collective excitations in the form of spin waves exists. The spin of the conduction electrons to which the magnetic impurity couples is determined not only by the spin of the particle-hole excitations in the leads, but also by the spin-waves. Therefore, the local moment is coupled to the fermionic bath of quasiparticles as well as the bosonic bath of spin-waves. For anti-aligned magnetization in the source and drain leads, the QPT of the Bose-Fermi Kondo model is predicted to take place as a function of gate voltage \cite{Kirchner2005_QCinFerro}. The gate voltage controls $\epsilon$, the energy of the localized state relative to the Fermi level, which determines the strength of the coupling to the fermionic and the bosonic bath. For $\epsilon$ comparable to the level broadening, the usual Kondo effect will take place.  As this energy ($\epsilon$) is increased, interactions with the spin-waves destroy the Kondo effect and drive the system to a universal local moment fluctuation phase. A QCP with non-Fermi liquid behavior is expected where the Kondo resonance is replaced by a fluctuating moment due to the coupling to the spin waves. Recent work has demonstrated the Kondo effect in a C-60 quantum dot with ferromagnetic leads \cite{Pasupathy2004_FeKondo}, as well as in a ferromagnetic atomic contact \cite{Calvo2009_KondoFeContacts}. These developments are a promising step toward realizing nanostructures that can model the Bose-Fermi Kondo effect.

%\bibliographystyle{nature}
%\bibliography{main2library}

\newcommand{\noopsort}[1]{} \newcommand{\printfirst}[2]{#1}
  \newcommand{\singleletter}[1]{#1} \newcommand{\switchargs}[2]{#2#1}

\end{document}